# Silicon-Organic Hybrid (SOH) Mach-Zehnder Modulators for 100 Gbit/s On-Off Keying

Stefan Wolf[1,†], Heiner Zwickel[1,†], Wladislaw Hartmann[1,2,3,†],
Matthias Lauermann[1,4], Yasar Kutuvantavida[1,2], Clemens Kieninger[1,2],
Lars Altenhain[5], Rolf Schmid[5], Jingdong Luo[6], Alex K.-Y. Jen[6], Sebastian Randel[1],
Wolfgang Freude[1], and Christian Koos[1,2,*]

[1]Institute of Photonics and Quantum Electronics (IPQ), Karlsruhe Institute of Technology (KIT), 76131 Karlsruhe, Germany
[2]Institute of Microstructure Technology (IMT), Karlsruhe Institute of Technology (KIT), 76131 Karlsruhe, Germany
[3]Now with: Physikalisches Institut, University of Muenster, 48149 Muenster, Germany
[4]Now with: Infinera Corporation, Sunnyvale, CA, United States
[5]Micram Microelectronic GmbH, 44801 Bochum, Germany
[6]Department of Material Science and Engineering, University of Washington, Seattle, Washington 98195, USA

†these authors contributed equally to the work

*email: christian.koos@kit.edu

**Contact Information for Authors:**

**Stefan Wolf**
s.wolf@kit.edu

**Heiner Zwickel**
heiner.zwickel@kit.edu

**Wladislaw Hartmann**
wladick.hartmann@uni-muenster.de

**Matthias Lauermann**
m.lauermann@posteo.de

**Yasar Kutuvantavida**
yasar.kutuvantavida@kit.edu

**Clemens Kieninger**
clemens.kieninger@kit.edu

**Lars Altenhain**
lars.altenhain@micram.com

**Rolf Schmid**
rolf.schmid@micram.com

**Jingdong Luo**
Jingdong.luo@soluxra.com

**Alex K.-Y. Jen**
ajen@u.washington.edu

**Sebastian Randel**
sebastian.randel@kit.edu

**Wolfgang Freude**
wolfgang.freude@kit.edu

**Corresponding Author:**
**Prof. Dr. Christian Koos**
Engesserstr. 5
76131 Karlsruhe, Germany

christian.koos@kit.edu

Phone:             +49 721 608-42491
Phone, secretary:  +49 721 608-42481
Fax, secretary:    +49 721 608-42786

**ABSTRACT**

**Electro-optic modulators for high-speed on-off keying (OOK) are key components of short- and medium-reach interconnects in data-center networks. Besides small footprint and cost-efficient large-scale production, small drive voltages and ultra-low power consumption are of paramount importance for such devices. Here we demonstrate that the concept of silicon-organic hybrid (SOH) integration is perfectly suited for meeting these challenges. The approach combines the unique processing advantages of large-scale silicon photonics with unrivalled electro-optic (EO) coefficients obtained by molecular engineering of organic materials. In our proof-of-concept experiments, we demonstrate generation and transmission of OOK signals with line rates of up to 100 Gbit/s using a 1.1 mm-long SOH Mach-Zehnder modulator (MZM) which features a π-voltage of only 0.9 V. This experiment represents not only the first demonstration of 100 Gbit/s OOK on the silicon photonic platform, but also leads to the lowest drive voltage and energy consumption ever demonstrated at this data rate for a semiconductor-based device. We support our experimental results by a theoretical analysis and show that the nonlinear transfer characteristic of the MZM can be exploited to overcome bandwidth limitations of the modulator and of the electric driver circuitry. The devices are fabricated in a commercial silicon photonics line and can hence be combined with the full portfolio of standard silicon photonic devices. We expect that high-speed power-efficient SOH modulators may have transformative impact on short-reach optical networks, enabling compact transceivers with unprecedented energy efficiency that will be at the heart of future Ethernet interfaces at Tbit/s data rates.**

**Keywords:** silicon photonics, electro-optic modulators, nonlinear optical devices, photonic integration, photonic integrated circuits (PIC), organic electro-optic materials, hybrid integration, energy-efficient communications

**INTRODUCTION**

Global data traffic continues to grow at double-digit annual rates[1], driven by cloud-based service delivery, video on demand, or Internet-of-Things (IoT) applications. To keep pace with this evolution, transceivers on all levels of optical networks are subject to the same challenge: To radically increase data rates, while maintaining acceptable technical complexity and energy consumption. In long-reach core and metropolitan networks, this challenge can be met by advanced modulation formats that exploit advances in high-speed digital signal processing (DSP) and advanced photonic integration. These approaches have led to transmission demonstrations at data rates in excess of 500 Gbit/s on a single polarization, using, e.g., 64-state quadrature amplitude modulation (64QAM) in conjunction with symbol rates up to 100 GBd[2,3]. When it comes to short-reach transmission over distances of a few kilometers or less in data centers or campus-area networks, however, higher-order modulation formats and the associated coherent reception techniques are prohibitive in terms of technical complexity and power consumption of the DSP. As a consequence, short-reach transmission largely relies on simple on-off-keying (OOK) as a modulation format that can be directly detected with a single high-speed photodiode. Such schemes are at the heart of current transceivers, e.g., for 100 Gbit/s Ethernet interfaces, that usually exploit low-cost vertical-cavity surface emitting lasers (VCSEL) or silicon photonic modulators to generate four spatially or spectrally separated data streams of 25 Gbit/s[4–6]. For future interfaces operating at data rates of 400 Gbit/s, 800 Gbit/s or 1.6 Tbit/s, however, parallelization of 25 Gbit/s channels is not a sustainable option[7]. Instead, lane rates of 100 Gbit/s are considered indispensable to maintain further scalability of throughput in campus-area and data-center networks. In this context, the key challenge is to build transmitters that can generate 100 Gbit/s OOK data streams at lowest possible power consumption, and such transceivers have even been identified as the "Holy Grail" of the Ethernet ecosystem by the Ethernet Alliance[7]. In fact, while a wide variety of compact high-speed electro-optic modulators has been demonstrated over the last years[8–16], only a few[10,11,16] are at all capable of providing OOK data rates of 100 Gbit/s, and these feature comparatively high operating voltages. The most efficient 100 Gbit/s devices demonstrated so far are InP Mach-Zehnder modulators (MZM) with an overall-length of 3 mm (or more) and featuring a π-voltage of 2 V[10,11].

In this paper we show that these limitations can be overcome by silicon-organic hybrid (SOH) modulators[17–31] that combine the advantages of large-scale silicon photonic integration with the extraordinarily high electro-optic (EO) coefficients obtained by molecular engineering of organic materials[32,33]. We demonstrate generation of OOK signals up to 100 Gbit/s using a 1.1 mm-long SOH MZM which features a π-voltage of only 0.9 V. The associated switching energy amounts to less than 98 fJ/bit – a record-low value for transmission at 100 Gbit/s

OOK using semiconductor-based modulators. In our experiments, we operate the device at a peak-to-peak voltage of 1.4 V, thereby exploiting the nonlinear transfer characteristic of the MZM to mitigate impairments by limited bandwidth of the modulator and of the transceiver circuitry, which leads to an estimated BER of down to $6.6 \times 10^{-6}$. We support our experiments by a theoretical analysis that takes into account the measured transfer function of the modulator and of the driver circuits. Besides signal generation, we transmit a 100 Gbit/s OOK stream over a dispersion-compensated 10 km standard single-mode fiber (SMF) link. This is the first transmission experiment of a 100 Gbit/s OOK signal generated by a semiconductor-based modulator. Unlike previous implementations[18,19] and competing device concepts[16,34–36] that rely on high-resolution electron beam lithography, our modulators were fabricated in a commercial silicon photonics line together with the full portfolio of silicon photonic devices and Ge photodiodes using standard 248 nm deep-UV lithography. In contrast to earlier demonstrations of SOH electro-optic modulators[18,19,24,30], this work focusses on high-speed serial transmission using simple on-off-keying and direct detection techniques rather than technically demanding quadrature amplitude modulation (QAM), thereby considerably reducing hardware complexity and avoiding energy-intensive signal processing. When combined with highly efficient CMOS drivers, SOH modulators have the potential to open a technically and commercially superior avenue towards short-reach transceivers with unprecedented energy efficiency that will be key for future Ethernet interfaces at Tbit/s data rates.

**MATERIALS AND METHODS**

**Principle of a silicon-organic hybrid (SOH) electro-optic (EO) phase modulator**

SOH modulators combine silicon-on-insulator (SOI) slot waveguides and electro-optic (EO) cladding materials[17]. A schematic of an SOH Mach-Zehnder modulator (MZM) and its cross-section are depicted in Figure 1 (a) and (b), respectively. The phase shifters consist of a silicon slot waveguide which is formed by two silicon rails[17]. Due to the discontinuity of the normal electric field component at the interface to the silicon rails, the dominant field component $\underline{\mathcal{E}}_{0,x}$ for the quasi-TE polarization in the slot is strongly enhanced, Figure 1 (b), Inset (1). The slot is filled with an organic EO material which provides a high $\chi^{(2)}$-nonlinearity (Pockels effect). The silicon rails are connected to aluminum (Al) electrodes via thin $n$-doped silicon slabs and aluminum vias (not depicted). A modulation voltage applied to the Al electrodes drops completely across the narrow slot and leads to a strong electric field $E_{x,RF}$ which is well confined to the silicon slot region, Figure 1 (b), Inset (2). The strong electric field interacts with the EO organic cladding and leads to a pronounced change of the refractive index, and consequently to a phase modulation of the optical wave. For estimating the phase shift $\Delta\Phi$ in an SOH waveguide, we may assume that the modulating radio-frequency RF field features only an $x$-component which has a constant value of $E_{x,RF} = U_{drive}/w_{slot}$ in the slot region $A_{slot}$, see Figure 1 (b), and which is negligible outside. This leads to the relation[17]

$$\Delta\Phi = \frac{1}{2} n_{EO}^3 r_{33} E_{x,RF} \Gamma k_0 L, \qquad (1)$$

where $k_0 = 2\pi/\lambda$ is the optical wavenumber at a vacuum wavelength $\lambda$, $\Gamma$ is the field interaction factor, $n_{EO}$ is the refractive index of the organic cladding material in the slot if no voltage is applied, $r_{33}$ is the EO coefficient and $L$ describes the length of the phase shifter. The field interaction factor $\Gamma$ can be calculated in terms of the vectorial mode fields $\underline{\mathcal{E}}_0(x,y)$ and $\underline{\mathcal{H}}_0(x,y)$ of the fundamental waveguide modes[17],

$$\Gamma = \frac{c\varepsilon_0 n_{EO} \iint_{A_{slot}} |\underline{\mathcal{E}}_{0,x}(x,y)|^2 \, dxdy}{\iint \mathrm{Re}\{\underline{\mathcal{E}}_0(x,y) \times \underline{\mathcal{H}}_0^*(x,y)\} \mathbf{e}_z \, dxdy}, \qquad (2)$$

where $n_{EO}$ is the refractive index of the organic cladding material in the slot, $c$ is the vacuum speed of light, and $\varepsilon_0$ is the vacuum dielectric constant.

The dynamic behavior of an SOH device can be understood by considering a lumped-circuit model as indicated in Figure 1 (b). It consists of a capacitor $C$ representing the SOH slot waveguide, and of two resistors $R$ which describe the finite conductivity of the $n$-doped silicon slabs. The resulting $RC$ low-pass characteristic as well as the RF propagation loss lead to bandwidth limitations[20]. The conductivity of the silicon slabs and hence the device bandwidth can be increased by, e.g., applying a voltage between the device layer and the bulk silicon, which leads to an electron accumulation layer and hence to a reduced resistivity of the slabs[37,20], see Figure 1 (b), Inset (3). Using this approach, bandwidths in excess of 100 GHz have been demonstrated[20]. The necessary gate voltage can be diminished by using thin oxide and doped poly-silicon gates deposited on top of the slab regions[26].

Note that the SOH approach maintains the full advantages of silicon photonics, exploiting highly mature CMOS processes for fabrication of the slot-waveguide base structures, onto which EO materials are deposited in a highly scalable post-processing step. In particular, SOH devices can be seamlessly integrated into complex photonic integrated circuits (PIC) that exploit the full range of devices available on the silicon photonic platform.

**Design, fabrication, and operation of SOH Mach-Zehnder modulators**

In this work we use a 1.1 mm long SOH Mach-Zehnder modulator (MZM) formed by two SOH phase shifter sections. The phase shifters comprise 240 nm wide rails and 160 nm wide slots and are covered by the organic EO material SEO100 [38], which has a refractive index of $n_{EO}$ = 1.73. This leads to a field interaction factor $\Gamma \approx 0.16$. Our experiments build upon a series of technological advances in comparison to earlier publications[19,22,24]. These advances relate, e.g., to fabrication techniques, to the materials, for which stability has been improved greatly[19], and to the underlying poling procedures. In particular, unlike previous generations of SOH modulators[18,19], the devices in this work are fabricated in a standard 248 nm deep ultra-violet (DUV) optical lithography process at A*Star IME, Singapore. The process allows co-integration with the full portfolio of silicon-photonic devices. The optical chip is globally covered with a thick layer of $SiO_2$ (not depicted in the figure), containing two metal layers for electrical wiring of the devices. To access the slot for deposition of the EO material, the oxide is locally removed by a dedicated backend process.

The electrodes of the MZM form a coplanar ground-signal-ground (GSG) transmission line, see Figure 1 (b). The EO material SEO100 is deposited onto the pre-processed device by spin-coating such that the slot is homogeneously filled. The material has a high EO coefficient of 166 pm/V as measured in bulk material for a wavelength of 1550 nm[39]. After spin coating, the macroscopic EO activity of the cladding is activated by poling[21] at an elevated temperature close to the material's glass transition temperature. To this end, a DC poling voltage is applied across the (floating) ground electrodes to align the EO chromophores in the two slots. The direction of alignment is defined by the direction of the electric DC poling field and is the same direction in both slots, indicated by green arrows in Figure 1 (b). The poling voltage remains applied while cooling the device to room temperature in order to freeze the chromophores in their state of orientation. Applying a modulation voltage to the signal electrode after poling leads to modulating fields oriented in opposite directions with respect to the chromophore alignment in the two slots. The modulating field is indicated by red arrows in Figure 1 (b). This leads to phase shifts of equal magnitudes but opposite signs in the two slots. This results in an efficient push-pull operation[17] and leads to chirp-free amplitude modulation provided that the device is perfectly balanced; see Supplementary Information for a more detailed discussion of chirp properties and of the impact of imbalance.

In general, SOH EO modulators stand out due to their high modulation efficiency, which can be expressed by the $\pi$-voltage-length product $U_\pi L$, where $U_\pi$ is the voltage required to achieve a phase difference of $\pi$ in the two arms of the MZM, and where $L$ denotes the phase shifter length. SOH MZM have been demonstrated with $U_\pi L$ products down to 0.5 Vmm[21,22] – more than an order of magnitude below that of conventional *pn*-depletion type devices[40,41]. Using SOH modulators, we have demonstrated optical signal generation with drive voltages down to 80 mV$_{pp}$ and energy consumptions of the order of 1 fJ/bit using OOK modulation[22]. In contrast to plasmonic-organic hybrid (POH) electro-optic modulators[34–36], which adapt the concept to plasmonic waveguides, SOH devices stand out due to significantly lower propagation losses, which enable larger lengths of phase shifters, and hence lower drive voltages[17]. As a quantitative measure, the product of the $\pi$-voltage $U_\pi$ and the achievable insertion loss $aL$ can be used, where $a$ denotes the propagation loss in the phase shifter in dB/mm and where $L$ is the phase shifter length. For POH modulators, this figure is usually above 10 dBV, whereas values of 1 dBV can be achieved by SOH devices[17]. The capabilities of the SOH platform can further be extended to highly efficient phase shifters based on liquid crystals[25], or to hybrid lasers that exploit light-emitting cladding materials[42].

While the present work concentrates on SOH MZM for high-speed OOK, the SOH modulator concept has also been proven to be perfectly suited for generation of advanced modulation formats such as 4-state pulse-amplitude modulation (4PAM, 120 Gbit/s) [28], 8-state amplitude shift keying[29] (8ASK, 84 Gbit/s), quadrature phase shift keying (QPSK), or 16-state quadrature amplitude modulation[18,23,24,30] (16QAM). For 16QAM signaling, we have recently demonstrated line rates of up to 400 Gbit/s [18,30]. At the same time, the efficiency of SOH modulators is accentuated by the ability for operation without external drive amplifiers, even for generation of higher-order modulation formats[23,24], where an electrical energy consumption of down to 18 fJ/bit has been demonstrated for 16QAM signaling. While these modulation formats lead to larger spectral efficiency and consequently to higher data rates, they considerably increase the complexity of transmitter and receiver and are hence not well suited for short-reach transmission in data centers or campus-area networks. This is particularly true for coherent communications where the receiver requires a dedicated photonic integrated circuit with two balanced detectors per polarization, a local oscillator laser, and extended DSP rather than just a simple photodiode.

The MZM used in our experiment features a π-voltage of 0.9 V at a wavelength of 1550 nm, see Figure 1 (c), which, for a device phase shifter length of 1.1 mm, corresponds to a $U_\pi L$-product of 1 Vmm. Note that the π-voltage was measured at DC bias voltages above 2 V. For smaller bias voltages, we observe slightly increased spacings of the transmission dips and hence slightly increased π-voltages, which is attributed to free ions in the cladding that lead to a partial screening of the applied fields at small bias voltages[17]. However, this effect is only observable for low frequencies and does not impede RF operation[21].

The slight increase of $U_\pi L$ in comparison to the value reported in references[21,22] is caused by a reduced $r_{33}$-coefficient of the presently used EO material, which was selected for high thermal stability rather than for highest EO activity. The EO coefficient $r_{33}$ can be estimated from the measured $U_\pi L$-product of the MZM operated in push-pull, the calculated field interaction factor $\Gamma$, and from the slot width $w_{slot}$,

$$U_\pi L = \frac{w_{slot}\lambda}{2n_{EO}^3 r_{33}\Gamma}, \qquad r_{33} = \frac{w_{slot}\lambda}{2U_\pi L n_{EO}^3 \Gamma} . \qquad (3)$$

In this relation, $\lambda$ is the carrier wavelength, and $n_{EO}$ is the refractive index of the organic cladding material[17]. The factor of 2 in the denominator results from the push-pull operation. For our device, we find an $r_{33}$ coefficient of 147 pm/V, which compares well to values of 166 pm/V reported for bulk SEO100, see reference [39]. The modulator structure as well as the organic material are well suited to operate over a large range of infrared telecommunication wavelengths comprising all relevant transmission bands between 1260 nm and 1675 nm, see Supplementary Information for details.

For the special SOH devices used in our experiments, rather high optical losses were observed with fiber-to-fiber attenuations of 20 dB or more. These high losses were caused by a fabrication problem, which lead to contamination of the slot waveguides with Germanium residuals – this problem was fixed in newer device generations by adapting the process flow. In the presented experiments, the losses of the devices comprise approximately 4.5 dB of fiber-chip coupling loss for each of the grating-coupler interfaces (9 dB in total), around 1 dB of excess loss for a pair of strip-to-slot converters, approximately 1 dB of excess loss for a pair of multi-mode interference couplers, and an additional approximately 1 dB for on-chip waveguides. This leaves approximately 8 dB for the 1.1 mm-long slot-waveguide section, corresponding to rather high propagation losses of 7.3 dB/mm for this specific device generation. These losses lead to $aU_\pi L$ products of 8 dBV, which is clearly above the 2.8 dBV that were previously demonstrated for SOH devices[22], but still well below the approximately 25 dBV found for POH modulators[43].

Another important performance parameter of MZM is the static extinction ratio $\delta^{(stat)}$, which is defined by the squared sum of the superimposed optical field strengths of both arms at the output of the MZM, related to the squared difference of the field strengths,

$$\delta^{(stat)} = \frac{(E_1+E_2)^2}{(E_1-E_2)^2}. \qquad (4)$$

If the splitting and combining ratios are not exactly 50/50 or if the loss in one arm is different from the loss in the other arm, then the amplitudes of the superimposed fields are different, and the extinction ratio (ER) is finite. This may lead to a chirped output signal, even if the modulator is operated in push-pull mode, see Section "Results and Discussion" and Supplementary Information for details. The (static) ER can be determined by measuring the transmission of the MZM as a function of the applied DC voltage, see Figure 1 (c). For the device used in the transmission experiments, the ER is rather low and amounts to approximately 14 dB. We attribute this to deviations of the MMI coupler from the ideal splitting ratio and to unequal propagation losses in the two arms of the MZM caused by the Germanium contaminations of the slot waveguides. Note that, due to the specific chip design, ports 2 and 4 of the MZM were inaccessible, see Figure 1 (a). Using ports 1 and 4 or ports 2 and 3 instead would eliminate the problem of non-ideal MMI couplers. In general, SOH devices can provide ER which are much better than the 14 dB obtained here – for other devices with similar device layouts we typically measure ER of $\delta^{(stat)}$ = (20 … 32) dB, see Supplementary Information and reference [21]. Note that the "dynamic" extinction ratio $\delta^{(dyn)}$ of the data signal is not only dictated by the static extinction ratio $\delta^{(stat)}$ of the modulator is modulator itself, but also by other effects such as inter-symbol interference, see Section "Generation and transmission of 100 Gbit/s OOK" below.

**RESULTS AND DISCUSSION**

**Setup for signal generation**

The experimental setup for data signal generation is depicted in Figure 2 (a). An external cavity laser (ECL) provides the optical carrier at a wavelength of approximately 1550 nm. The light is coupled to and from the SOH MZM via grating couplers. While these couplers are perfectly suited for testing, they introduce limitations of the operating wavelength range. In advanced device implementations, these grating couplers might be replaced by edge coupling, e.g., based on free-space assemblies of micro-lenses and prisms[44], or with 3D freeform waveguides or lenses printed by two-photon polymerization[45–49]. After modulation, the optical signal is fed into an erbium-doped fiber amplifier (EDFA) with a gain of 35 dB, an optional 10 km long fiber, and a 2 nm wide optical band-pass filter to remove out-of-band noise, before being detected by a 100 Gbit/s photodiode. At the transmitter, an arbitrary waveform generator (AWG, Keysight M8195A) is used to synthesize the electrical drive signals, using two independent pseudo-random binary sequences (PRBS) of length $2^9$-1. These signals enter a 2:1 electrical multiplexer (MUX, SHF 603A), the output of which is a non-return-to-zero (NRZ) signal with a peak-to-peak voltage swing of 0.4 $V_{pp}$ and double the symbol rate of the inputs. A radio-frequency (RF) amplifier (SHF 827) with a nominal bandwidth of 70 GHz is used to boost the signal at the MUX output to a peak-to-peak voltage of 1.4 $V_{pp}$. This signal is then coupled to the GSG transmission line of the MZM via microwave probes having a nominal bandwidth of 67 GHz. A DC bias voltage is applied via the same microwave probe to set the operating point of the MZM to the quadrature (3 dB) point. An external 50 Ω termination resistor connected to the end of the transmission line via a second microwave probe prevents back-reflections. Because de-multiplexers operating at 100 Gbit/s are commercially not yet available, we could not analyze the data stream in real-time, and it was hence impossible to measure the bit error ratio (BER) directly. Instead, the received electrical signal is analyzed using an Agilent 86100C digital communications analyzer (DCA) with a 70 GHz equivalent-time sampling module (Agilent 86118A). We record the electrical eye diagrams and extract the quality factor (Q-factor) which is defined by the signal's mean levels $u_1$ and $u_0$ for the logical '1' and the logical '0', and by the corresponding standard deviations $\sigma_1$ and $\sigma_0$,

$$Q = \frac{u_1 - u_0}{\sigma_1 + \sigma_0}. \tag{5}$$

The BER of the data signal can be estimated from the measured Q-factor[50]

$$\mathrm{BER}_e = \frac{1}{2}\mathrm{erfc}\left(\frac{Q}{\sqrt{2}}\right). \tag{6}$$

The complementary error function is defined by $\mathrm{erfc}(z) = \left(2/\sqrt{\pi}\right)\int_z^\infty \exp(-t^2)\mathrm{d}t$.

**Generation and transmission of 100 Gbit/s OOK**

The NRZ eye diagrams of the OOK drive signals for different data rates are depicted in Figure 2 (b), (c). The clearly open eyes of the MUX output are shown in the first column, Figure 2 (b), while the eye diagrams of the RF drive amplifier output are depicted in the second column, Figure 2 (c). Up to data rates of 60 Gbit/s, the eyes are well open, while at 80 Gbit/s the signal starts deteriorating. This is due to the bandwidth limitations of the amplifier, which predominantly affect its phase response and lead to significant group delay dispersion, see Section "Theoretical analysis of bandwidth limitations" for details.

In a back-to-back measurement without the 10 km long transmission fiber, we record the eye diagrams of the optical signals after direct detection of data signals from 60 Gbit/s to 100 Gbit/s, Figure 2 (d). For data rates of 60 Gbit/s and 70 Gbit/s, we measure open eyes and Q-factors of Q = 5.7 and 4.8, respectively. Following Eq. (6), we estimate $\mathrm{BER}_e$ of $6.7 \times 10^{-9}$ and $7.9 \times 10^{-7}$. At 80 Gbit/s and 100 Gbit/s, the measured Q-factors of 4.2 and 3.4 correspond to an estimated $\mathrm{BER}_e$ of $1.3 \times 10^{-5}$ and $3.2 \times 10^{-4}$, respectively. All of these $\mathrm{BER}_e$ values are below the threshold of $4.5 \times 10^{-3}$ for hard-decision forward error correction (FEC) with 7% overhead[51]. A gate field was applied for the 100 Gbit/s signal only. The 100 Gbit/s achieved in our experiment corresponds to the highest OOK data rate generated by a silicon-based modulator so far, see Section "Competitive benchmarking and application potential" for a more detailed comparison to other experiments. Note that in our as well as in competing high-speed OOK demonstrations[10,11,16], the data rate of 100 Gbit/s does not refer to the net data rate, but to the line rate and hence includes the 7 % FEC overhead.

Note that the rather low static extinction ratio (ER) of approximately 14 dB of our MZM leads to a residual chirp of the generated data signal, which would not occur for perfectly balanced devices. For quantifying the chirp of the data signal, we use the chirp parameter $\alpha$ that is essentially defined by the ratio of the phase modulation to the amplitude modulation[52,53],

$$\alpha = 2P \frac{\mathrm{d}\phi/\mathrm{d}t}{\mathrm{d}P/\mathrm{d}t}. \qquad (7)$$

In this relation, $\phi$ denotes the phase and $P$ the time-dependent power of the optical signal averaged over a few optical cycles. In SOH devices, imbalance of the MZM arms is the dominant source of chirp, see Supplementary Information for a more detailed discussion and an experimental verification. The magnitude of the chirp parameter $\alpha$ can thus be directly related to the ratio $\gamma = E_2/E_1$ of the fields in the two MZM arms[52]

$$|\alpha| = \frac{1}{\gamma} \frac{\phi_1 + \gamma^2 \phi_2}{\phi_1 - \phi_2}. \qquad (8)$$

In this relation, $\phi_1$ and $\phi_2$ denote the phases shifts in the individual MZM – for push-pull modulation we can assume $\phi_2 = -\phi_1$. The field amplitude ratio $\gamma$ can be derived from a measurement of the (static) extinction ratio $\delta^{(\mathrm{stat})}$ defined in Eq. (4),

$$\gamma = \frac{\sqrt{\delta} - 1}{\sqrt{\delta} + 1}. \qquad (9)$$

Note that the electric fields are chosen such that $E_1 > E_2$ and hence $0 \leq \gamma \leq 1$. As a consequence, Eq. (8) only allows determining the magnitude of $\alpha$, but not its sign – this would require knowing whether the stronger optical amplitude is associated with the MZM arm having a positive or a negative phase shift, which cannot be derived from a measurement of the static extinction ratio. Using Eqs. (8) and (9), the static extinction ratio of 14 dB of our device translates into a magnitude of the chirp parameter of $|\alpha| \approx 0.42$, which is well below chirp parameters of $|\alpha| = 0.8$ that are obtained for conventional *pn*-depletion type silicon modulators with comparable ER[54].

We also measure the dynamic extinction ratios of our data signals. For 100 Gbit/s signaling, the measured dynamic ER amounts to 5 … 7 dB, which compares well to a measured extinction ratio of 6.1 dB that was observed for a conventional *pn*-depletion type silicon modulator[14] at 70 Gbit/s. Note that the dynamic extinction ratio measured from the eye diagram of a data signal is generally worse than the static extinction ratio of the underlying modulator according to Eq. (4). This is due to inter-symbol interference and quadratically detected optical noise.

In addition to the back-to-back experiment, we transmit the data signals over a dispersion-compensated fiber link of 10 km standard single-mode fiber (SMF) having a negligible residual dispersion of 2.6 ps/nm. The eye diagrams of the data signals received after transmission are depicted in Figure 2 (e). For the 60 Gbit/s and 70 Gbit/s data, the Q-factors of 5.6 and 4.4 did not significantly deteriorate compared to the Q-factors of 5.7 and 4.8 in the back-to-back measurements. These Q-factors correspond to $\mathrm{BER_e} = 7.9 \times 10^{-7}$ and $\mathrm{BER_e} = 4.9 \times 10^{-6}$, respectively. For data rates of 80 Gbit/s and 100 Gbit/s, a gate field was applied and the measured Q-factors of 4.2 and 2.8 correspond to $\mathrm{BER_e}$ values of $1.5 \times 10^{-5}$ and $3.0 \times 10^{-3}$. This experiment corresponds to the first 100 Gbit/s OOK transmission demonstration using a modulator on the silicon photonic platform. The transmission demonstration over the dispersion-compensated 10 km fiber link was performed with binary drive signals without further signal processing. Still, the results compare very well to recently published demonstrations of 100 Gbit/s OOK transmission over an uncompensated 1.8 km-link using InP based devices in combination with digital equalization at the receiver[11].

**Theoretical analysis of bandwidth limitations**

To analyze the impact of bandwidth limitations on our experiments, we reproduce the results by simulations, see Figure 3. To this end, we use a vector network analyzer (VNA) to measure the frequency response of a 1.1 mm long SOH MZM with a gate field of 0.1 V/nm, as used for the 100 Gbit/s transmission experiment. The VNA generates an input signal with varying frequency, which is coupled to the (terminated) modulator. The modulated optical power is received with a photodiode featuring a calibrated frequency response, which feeds its output back to the VNA for a characterization of the electro-optic-electric (EOE) bandwidth. The measured frequency response (by modulus and phase) is depicted in Figure 3 (a). The 6 dB point[55] of the EOE frequency response is found to be about 25 GHz.

At first sight, a 6 dB EOE bandwidth of 25 GHz seems rather small for generating a 100 Gbit/s NRZ signal. In general, considering a low-pass filter at the receiver, an optimum signal-to-noise power ratio is found for a 3 dB bandwidth of approximately 65 % of the symbol rate[56]. This bandwidth leads to an ideal trade-off of noise power in the filter passband and inter-symbol interference (ISI) caused by the limited filter bandwidth. In our case, however, the situation is different: The bandwidth limitation is caused by a low-pass characteristic associated with the electrical part of the modulator, which is then followed by the nonlinear, cosine-shaped MZM intensity transmission characteristic, see Figure 3 (d). As a consequence, while low amplitudes of the drive signal translate linearly to optical output power, higher amplitudes are compressed, thereby mitigating ISI-related amplitude fluctuations of the drive signal. To illustrate this effect, we analyze our transmission system in MATLAB. To emulate the limited rise and fall times of the hardware in the simulation, we use cosine-shaped pulses in the time-domain (not to be confused with raised-cosine pulse shaping with a raised-cosine shaped spectrum) to approximate the measured output signal of the MUX, see Figure 2 (b). Details can be found in the Supplementary Information. In the simulation, this signal is then fed to the drive amplifier, which has a gain of 11 dB and which is modeled by the measured $S_{21}$-parameter, Figure 3 (b). At the output of the amplifier, the drive signal features a peak-to-peak voltage swing of 1.4 $V_{pp}$, which is used as an input signal to the modulator. The modulator is modeled by the measured frequency response depicted in Figure 3 (a), followed by the cosine-shape time-domain power transfer function, for which we assume a $\pi$-voltage of 0.9 V. The frequency response was set to nearly zero (-200 dB) beyond the measured frequency range. The results of the simulation are depicted in the lower part of Figure 3 (a). Interestingly, the electrical eye diagrams at the amplifier output show already significant distortions, see left column of eye diagrams in Figure 3 (a), even though the magnitude of the frequency response of the device remains flat up to more than 70 GHz. It turns out that these distortions are not caused by the low-pass characteristic of the amplifier, but by its phase response, which starts dropping by approximately -200 ° between 40 GHz and 75 GHz, see Figure 3 (b). This can be confirmed by assuming a flat phase characteristic in the simulation, for which the distortions disappear. The low-pass modulator frequency response leads to further closure of the eye due to ISI, see center column of eye diagrams in Figure 3 (a), and it is only the compression of large amplitudes in the MZM that reproduces an open eye, see right column of eye diagrams in Figure **3** (a). Using this property allows the generation of high-speed NRZ signals with data rates well beyond the modulator's small-signal 6 dB bandwidth. Note that, in contrast to other 100 Gbit/s OOK demonstrations[11,16] or high symbol-rate coherent modulation experiments[30], our demonstration does not require any digital pre-distortion or post-equalization and can rely on simple binary drive signals for the modulator at the transmitter and on a simple sampling oscilloscope for measuring the eye diagram at the receiver.

Note that the compression of large amplitudes by the cosine-shaped transfer function is a general property of all MZM. The effect, however, can only be exploited if the available peak-to-peak drive voltage reaches the $\pi$-voltage of the device. It is a unique feature of our SOH MZM that this amplitude compression can be achieved at comparatively low drive voltages, which can be realistically generated by currently available driver circuits. The low operating voltage becomes particularly crucial at high data rates, where drive signals with high modulation amplitude are particularly difficult to generate. As an example, conventional *pn*-depletion-type modulators require peak-to-peak drive voltage swings in excess of 5 V to reach compression of the power transfer function, which would be hard to achieve with currently available electrical drivers for 100 Gbit/s, in particular when lower power consumption shall be maintained.

In our experiments, the length of the PRBS was limited to $2^9 - 1$ due to the memory size and the memory granularity of the AWG, but does not represent a fundamental limit of the device. Using a longer PRBS sequence would lead to elongated sequences of subsequent logical "0" or "1" and hence to more low-frequency components in the signal spectrum. This should not affect the results – the lower cut-off frequency of our transmitter amounts to only 70 kHz, dictated by the AC-coupled drive amplifier. To confirm this expectation, we simulated PRBS sequences of lengths $2^9 - 1$ and $2^{18} - 1$ – the maximum possible length that could be handled by our computer. We do not find any relevant change of the received eye diagram.

**Signal generation using a 100 GSa/s digital-to-analog converter (DAC) and a real-time oscilloscope**

With our previous experiments we were unable to measure the BER directly, since the eye diagram at the receiver was taken with equivalent-time sampling technique. For a more direct analysis of the BER, we improved our setup as follows: AWG and MUX were replaced by a programmable DAC with a sampling rate of 100 GSa/s (MICRAM DAC4) and analog bandwidth above 40 GHz, and the equivalent-time sampling oscilloscope was substituted by a real-time oscilloscope having a bandwidth of 63 GHz. For detection, we use a photodiode with a bandwidth of 70 GHz, similar to the one of our previous experiment. We record the time traces of the photodiode current and analyze it offline with MATLAB. We operate the MZM without a gate field. For a 100 Gbit/s OOK signal we measure a Q-factor of 2.4 and estimate a $BER_e = 8.2 \times 10^{-3}$, Figure 4 (a). The Q-factor in the previous experiment was larger and the estimated $BER_e$ was smaller because the MUX acted as a limiter, and because the gate field increased the MZM bandwidth. When taking the directly measured BER from the DAC-based experiment, however, we find a value of only $BER_m = 4.2 \times 10^{-3}$, which is smaller than the $BER_e$ estimated from the measured Q-factor and falls just within the limits of hard-decision (HD) FEC with 7% overhead. This illustrates the issues involved in estimating a BER from the Q-factor using Eq. (6).

Note that FEC is presently not used for short-reach optical interconnects because of the computational latency. However, the OpenOptics multi-source agreement (MSA)[57] already describes the use of optimized, low-latency FEC codes that allow to trade coding gain for latency[58] and therefore exhibit more stringent BER thresholds. With a post-equalization filter, we measure a $BER_m$ of $6.6 \times 10^{-6}$, see Figure 4 (c), which is within the limits of low-latency FEC. On-chip equalizers are commonly used in state-of-the-art transmitter drive chips[59].

**Energy considerations**

In data center networks, power dissipation is a severe aspect, and the driver electronics of EO modulators play an important role in the overall energy consumption of the transceiver. The design and therefore the power dissipation of the driver depends strongly on the specifications of the modulator. In the following, we analyze the modulator's energy consumption per bit, which we regard as a figure of merit for the overall power dissipation of the transceiver. To this end, we assume that the GSG transmission line impedance is matched to the 50 Ω drive circuitry and to the terminating resistor $R_L$ = 50 Ω, see Figure 2 (a). The modulator is driven with rectangular non-return-to-zero (NRZ) pulse sequences, for which logical "1" and "0" are equiprobable. The energy consumption per bit can then be calculated[22] by dividing the electrical power associated with the drive voltage amplitude by the line rate $r$,

$$W_{bit} = \left(\frac{U_{drive}}{2}\right)^2 \frac{1}{R_L r}. \qquad (10)$$

For $R_L$ = 50 Ω, a peak-to-peak voltage swing of $U_{drive}$ = 1.4 V, and a line rate of 100 Gbit/s, we find an energy consumption of 98 fJ/bit. This is, to the best of our knowledge, the smallest power consumption ever reported for a semiconductor-based MZM at 100 Gbit/s OOK. Note that the biasing and the gate voltage do not involve any measureable DC current flow and hence do not contribute to the energy consumption.

**Competitive benchmarking and application potential**

The experiments presented in the previous sections demonstrate the unprecedented performance of SOH EO modulators to realize high-speed energy efficient OOK transceivers for data-center and campus area networks. A more detailed comparison to competing concepts of semiconductor-based devices is given in Figure 5 (a). The graph shows the required peak-to-peak drive voltages as a measure of the energy consumption in dependence of experimentally demonstrated OOK line rates. Energy-efficient and fast modulators are found in the lower right corner of the diagram. With our **SOH MZM**, we generate line rates of up to 100 Gbit/s at drive voltages as low as 1.4 $V_{pp}$ (★), corresponding to a switching energy of 98 fJ/bit. This marks the highest OOK line rate reported so far at the lowest drive voltage. A previous OOK data transmission[22] experiment at 40 Gbit/s with a drive voltage of only 950 mV is also given as a reference.

The fastest and most energy-efficient competing devices are based on InP, where both electro-absorption modulators (EAM) and MZM have been shown. **InP EAM** enable line rates up to 80 Gbit/s at peak-to-peak drive voltages of 3 $V_{pp}$[8]. With an **InP MZM**, 60 Gbit/s OOK signaling was demonstrated at a comparatively low drive voltage of 1.5 V[9]. More recently 100 Gbit/s at drive voltages of 2.3 $V_{pp}$ were shown[10]. This is clearly higher than the 1.4 $V_{pp}$ demonstrated for our devices, and the associated switching energy of more than 250 fJ/bit is more than twice the value reported for our device. Further, 100 Gbit/s OOK was demonstrated[11] with an InP MZM at a peak-to-peak drive voltage of 1.5 $V_{pp}$. This performance is on par with the results obtained by our experiment, however the device features a π-voltage of 2 V which is twice as high as the 0.9 V obtained for the SOH MZM. This allowed to drive the modulator in the linear regime for multi-level PAM signaling but sacrificed modulation depth. Moreover, SOH devices exploit the intrinsic scalability advantages associated with large-scale silicon photonic fabrication based on highly developed CMOS processes, potentially in combination with electrical circuits[60,61]. Regarding the silicon photonic platform, silicon-germanium (**SiGe**) **EAM** were realized and demonstrated at line rates of up to 50 Gbit/s, using drive voltages of at least 2 $V_{pp}$[12,13]. With all-silicon MZM (**Si MZM**), line rates up to 70 Gbit/s[14] and 80 Gbit/s[15] were demonstrated, but at the expense of large drive voltages of 5 $V_{pp}$ and beyond. Another interesting approach published recently relies on a silicon-based plasmonic-organic hybrid (**POH**) **MZM**, for which a line rate of 72 Gbit/s and 100 Gbit/s have been published[16]. It shall be noted that the 72 Gbit/s signal was detected by a coherent receiver and hence signal quality aspects cannot be directly compared to the other direct detection experiments. In general, the performance of these devices is fundamentally limited by an intrinsic trade-off between energy consumption and insertion loss[17], leading to relatively large drive voltages of 6 $V_{pp}$[36] and 4 $V_{pp}$[16], respectively. In addition, fabrication of POH devices still requires advanced electron-beam lithography, which is not amenable to large-scale low-cost production. Figure 5 (b) takes the same MZM references but plots the π-voltages instead of the applied drive voltage. The plot nicely exhibits the extraordinary low π-voltages of less than 1 V as the unique feature of SOH modulators. For the **InP MZM**, the π-voltages amount to 1.7 V and 2 V for the devices demonstrating 60 Gbit/s and 100 Gbit/s, respectively. The π-voltage of the **POH MZM** is not specified in the paper, but estimated from the device length and from the $U_\pi L$ product that is referenced for POH modulators using the same EO material. Note that the EAM listed in Figure 5 (a) are not included in the comparison in (b), simply because the π-voltage is not defined for the EAM. Moreover, for some of the MZM devices such as the 70 Gbit/s **Si MZM**[14] and the 72 Gbit/s **POH MZM**[36], the π-voltages of 20 V and around 10 V are simply outside the chart. For VCSEL-based links, the highest OOK data rate demonstrated so far amounts to 70 Gbit/s [5,62], but transmission distances are usually less than 100 m.

SOH devices can hence compare very well to competing device concepts, both in terms of performance and in terms of device scalability. The record-low drive voltage of 1.4 $V_{pp}$ can be further reduced by using optimized poling techniques[63] or advanced organic materials, for which EO coefficients in excess of 350 pm/V have been demonstrated[31]. This might pave the path towards SOH MZM with sub-500 mV drive voltages that can be directly operated[23] by energy-efficient CMOS circuits without additional SiGe BiCMOS amplifier stages. Note that serializer/deserializer (SerDes) chips generating 100 Gbit/s NRZ drive signals have recently been realized on the BiCMOS platform[59]. SOH modulators are perfectly suited to complement these circuits on the optical side and to enable short-reach high-speed transceivers with unprecedented energy efficiency that will be key for future Ethernet interfaces at data rates of 400 Gbit/s, 800 Gbit/s, or 1.6 Tbit/s. In the future, such devices might be co-integrated with light sources in compact chip-scale assemblies. This can be accomplished by hybrid integration approaches that rely, e.g., on flip-chip integration of direct-bandgap III-V dies on processed silicon photonic waveguides[64], on mounting of readily processed III-V lasers onto silicon photonic dies[65], or on photonic multi-chip integration concepts that exploit the concept of photonic wire bonding[45,47,48].

Another important aspect of the SOH concept is the stability of the organic EO materials. Recent technological advances in material synthesis and molecular design have led to efficient and thermally stable organic EO materials. Previous experiments for record-low energy consumption[22,66] used materials that have been optimized for electro-optic activity but lacked temporal stability. Meanwhile, device tests at elevated temperatures of 85 °C have shown stable operation for 500 hours for SEO100 [38] and other materials[67], and further progress is to be expected. In a different set of experiments, we showed that SOH modulators with SEO100 are suited for operation at elevated temperatures of 80°C under ambient atmospheric conditions[19]. In the future, even higher operating temperatures are expected, e.g., by cross-linking techniques[68] or by using materials with intrinsically higher glass transition temperatures such as side-chain EO polymers[67]. In addition to temperature, photo-oxidation might play an important role in the degradation of EO organic materials – this aspect is subject to ongoing research, and we expect that the associated life-time limitations can also be overcome by cross-linking, which reduces oxygen diffusion into the material[69,70], or by encapsulation of the devices.

**CONCLUSION**

Using a silicon-organic hybrid (SOH) MZM we demonstrate for the first time the generation and transmission of a 100 Gbit/s OOK signal at record-low drive voltages of only 1.4 $V_{pp}$ and energy consumptions of only 98 fJ/bit. We confirm our experimental results by simulations using the measured frequency characteristic of our SOH modulator and its associated drive circuitry. The moderate EOE modulator bandwidth of 25 GHz leads to an eye opening, which is much improved by the nonlinear modulator transfer characteristic. In the experiments, we achieve BER below the threshold for hard-decision forward error correction. Post-equalization helps in improving the eye opening so that low-power and low-latency FEC codes can be employed. The efficiency of our modulator and the possibility to exploit large-scale silicon photonic integration allows the realization of compact and technically simple high-speed transceivers that meet the stringent cost targets of medium-reach interconnects.

**ACKNOWLEDGEMENTS**

We thank SHF Communication Technologies AG for lending the multiplexer. We acknowledge support by the European Research Council (ERC Starting Grant 'EnTeraPIC', 280145), by the EU FP7 projects PhoxTroT and BigPipes, by the Alfried Krupp von Bohlen und Halbach Foundation, by the Helmholtz International Research School for Teratronics (HIRST), by the Karlsruhe School of Optics and Photonics (KSOP), and by the Karlsruhe Nano-Micro Facility (KNMF), and by Air Force Research Laboratory (AFRL) and Air Force Office of Scientific Research (AFOSR) through the contract of FA8650-14-C-5005 under the Small Business Technology Transfer Research program (STTR). We acknowledge the support from Keysight Technologies in Boeblingen, Germany, for lending the real-time oscilloscope. Furthermore, we acknowledge support by Deutsche Forschungsgemeinschaft and Open Access Publishing Fund of Karlsruhe Institute of Technology.

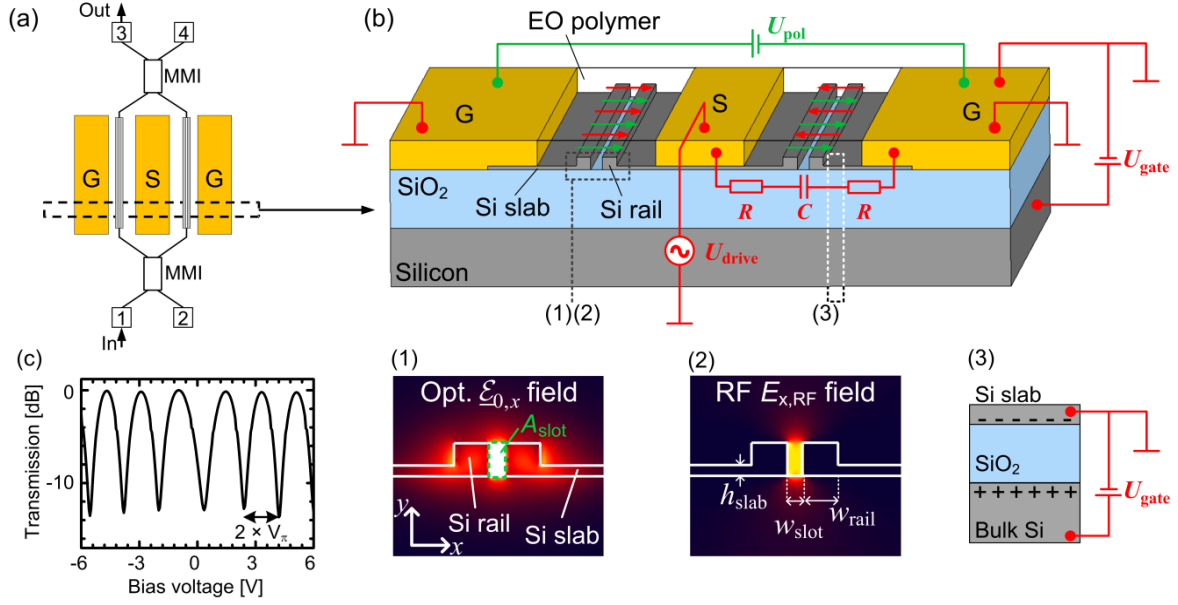

**Figure 1:** Schematic of an SOH Mach-Zehnder modulator (MZM). **(a)** Top view of an MZM with two SOH phase shifters, a coplanar ground-signal-ground (GSG) transmission line, and a pair of multi-mode interference (MMI) couplers. **(b)** Cross-sectional view of the phase shifter section of the SOH MZM fabricated on a silicon-on-insulator substrate with a 2 µm thick buried oxide (BOX) layer. The slot waveguide ($w_{Slot}$ = 160 nm, $w_{Rail}$ = 240 nm, $h_{Rail}$ = 70 nm) is formed by two silicon rails and is embedded into an organic electro-optic (EO) cladding material. The slot is connected to the aluminum (Al) transmission lines by thin *n*-doped silicon slabs and aluminum vias (not depicted). A poling voltage $U_{pol}$ applied across the (floating) ground electrodes at an elevated temperature close to the material's glass transition point aligns the EO chromophores in the slot (green arrows). An electric field generated from a modulation signal $U_{drive}$ applied to the GSG electrodes (red arrows) is oriented parallel (anti-parallel) with respect to the chromophores orientation in the left (right) slots. This leads to a phase shift of equal magnitude but opposite sign in the two slots, resulting in chirp-free push-pull operation. The electronic bandwidth of an SOH MZM is limited by the inherent *RC* lowpass characteristic resulting from the limited conductivity of the *n*-doped silicon slabs and the capacitance of the slot. Inset **(1)**: Dominant *x*-component of the optical electric field $E_{x,opt}$ in the slot waveguide with slot area $A_{slot}$. Inset **(2)**: Field component $\mathcal{E}_{0,x}$ of the electrical RF drive signal. Both the optical and the electrical field are well confined to the slot and overlap strongly for an efficient modulation. Inset **(3)**: Electron accumulation layer. The modulator bandwidth is increased by a decreased resistance (*R*) of the slabs, induced by a charge accumulation layer that can be generated by a "gate voltage" $U_{gate}$ between the bulk silicon and the ground electrodes. **(c)** DC characteristic of a 1.1 mm long SOH MZM with a $\pi$-voltage of 0.9 V measured at a voltage offset slightly more than 2 V. Using this offset avoids screening effects of the applied electric field by free charges in the organic cladding [17].

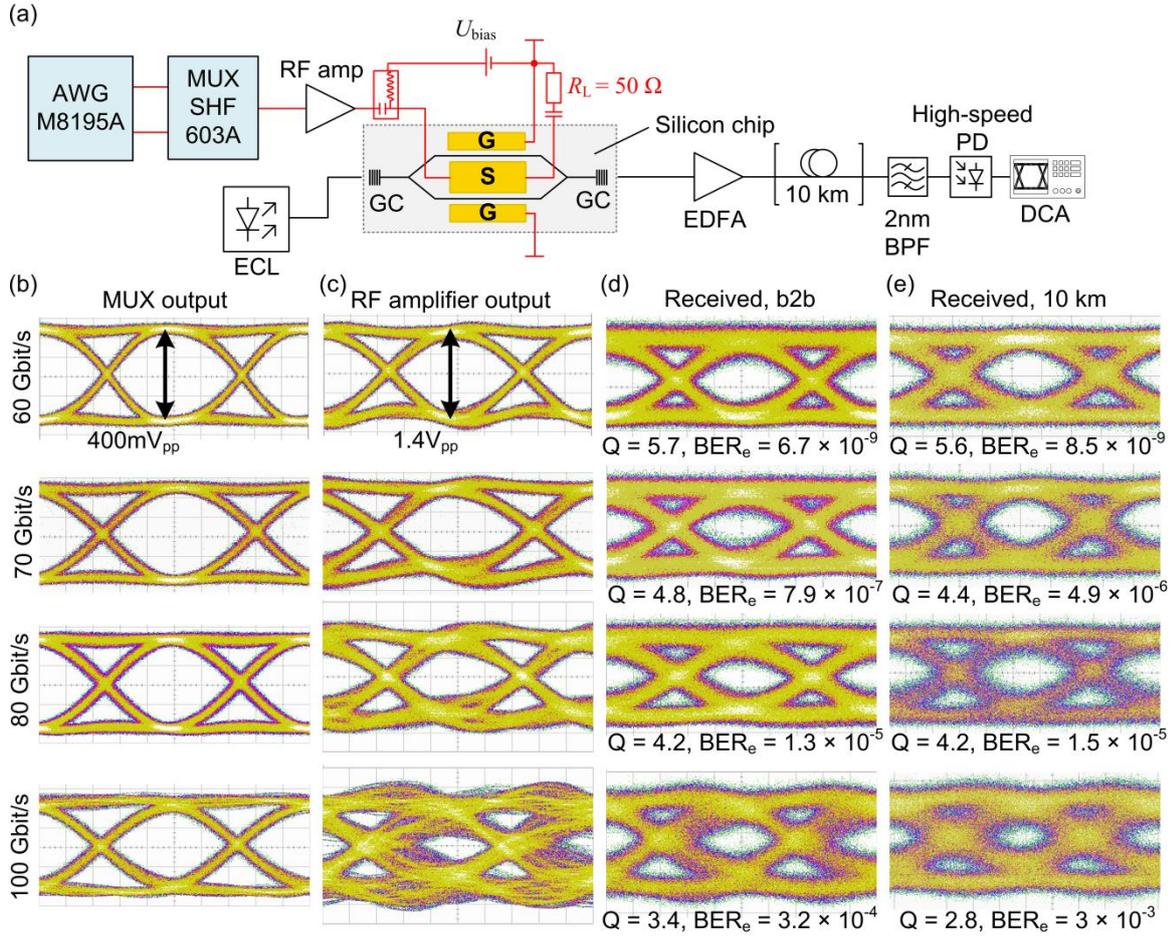

**Figure 2**: Experimental setup and measured eye diagrams. **(a)** Setup for 100 Gbit/s OOK data generation. An external cavity laser (ECL) provides the optical carrier. Optical power is coupled to and off the 1.1 mm long SOH MZM chip via grating couplers (GC). An erbium-doped fiber amplifier (EDFA) with a 2 nm optical bandpass filter (BPF) compensates chip losses. An optional dispersion-compensated 10 km fiber link is used for transmission experiments. A 100 Gbit/s photodiode detects the signal and feeds it to a 70 GHz equivalent-time sampling module of a digital communications analyzer (DCA). The electrical drive signal is derived from two independent pseudo-random binary sequences (PRBS) generated in an arbitrary waveform generator (AWG, Keysight M8195A). These sequences are fed to a multiplexer (MUX, SHF 603A), combined to a single binary NRZ sequence (≤ 100 Gbit/s), amplified with a radio-frequency (RF) amplifier, and coupled to the chip using microwave probes. A bias-T adds a DC voltage $U_{bias}$ to set the MZM to the quadrature (3 dB) operating point. The ground-signal-ground (GSG) transmission line is terminated with a 50 Ω resistor. **(b)** Eye diagrams of MUX output (voltage swing 400 mV$_{pp}$) for data rates 60 Gbit/s, 70 Gbit/s, 80 Gbit/s, and 100 Gbit/s. **(c)** Eye diagrams of RF amplifier output (voltage swing 1.4 V$_{pp}$). The amplifier transfer function causes signal distortions. **(d)** Eye diagrams, measured Q-factor and estimated BER$_e$ after detection (back-to-back, b2b) and **(e)** after transmission over the dispersion-compensated 10 km long link. In the b2b measurement, a gate field was applied for the 100 Gbit/s signal only. For the transmission, a gate field was applied for both the 80 Gbit/s and 100 Gbit/s signal.

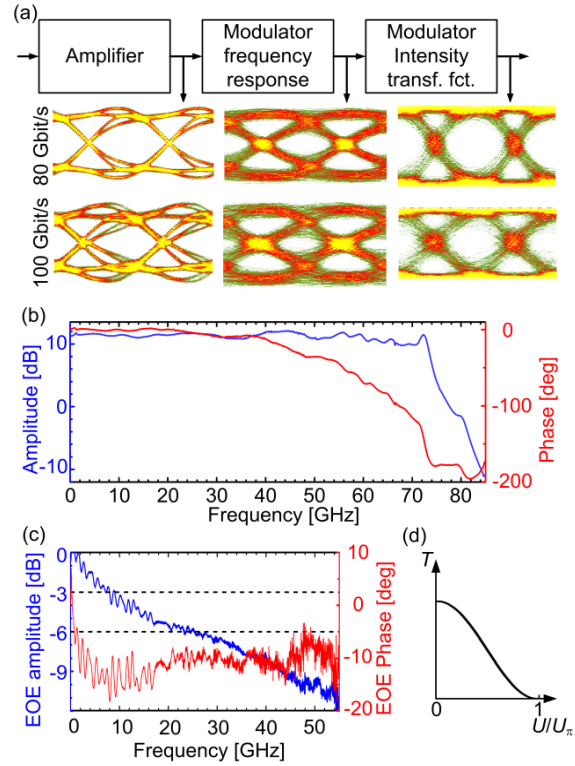

**Figure 3:** Analysis of bandwidth limitations. **(a)** Simulated eye diagrams. Amplifier input: Cosine-shaped pulses with full-width-half-maximum (FWHM) duration equal to the symbol period. The amplifier is modeled by its measured frequency response, see (b). The modulator is represented by its measured small-signal transfer function, see (c), followed by a co-sine-shaped intensity transfer characteristic as sketched in (d). Color map differs from Figure 2. **(b)** Modulus and phase of measured $\underline{S}_{21}$ parameters of 70 GHz drive amplifier. **(c)** Modulus and phase of measured small-signal electro-optic frequency response of a 1.1 mm long SOH MZM. Note that the bandwidth of an EO modulator is usually specified by the modulation frequency which corresponds to a 6 dB drop of EOE response [10,55]. For the spectral component associated with the modulation frequency, this corresponds to a drop of the optical power and hence of the photocurrent amplitude by factor of two, which is measured as a four-fold (6 dB) decrease of the spectral power density by the VNA. For our device, the 6 dB bandwidth amounts 25 GHz. Organic cladding: SEO100. Gate field: 0.1 V/nm. **(d)** Intensity transfer function $T$ of a MZM vs. normalized drive voltage $U/U_\pi$.

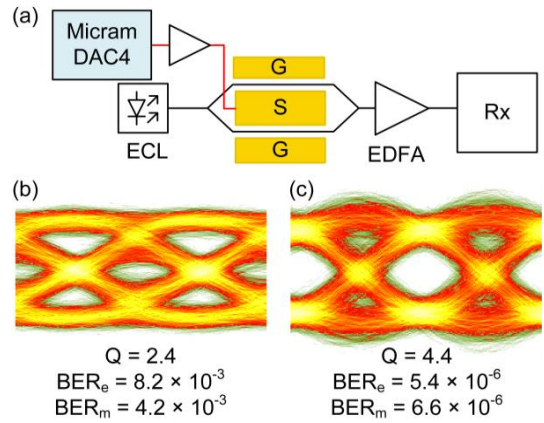

**Figure 4:** Experimental b2b setup and measured eye diagrams for a 100 Gbit/s OOK. (a) A programmable digital-to-analog converter (DAC, 100 GSa/s) serves as an AWG and drives the MZM (no gate field). The receiver Rx comprises a BPF, a photodiode, and a 63 GHz real-time oscilloscope. (b) Received eye diagram, measured Q-factor, estimated $BER_e$, and measured $BER_m$ at 100 Gbit/s without post-equalization and (c) with post-equalization. Color map differs from Fig. 2.

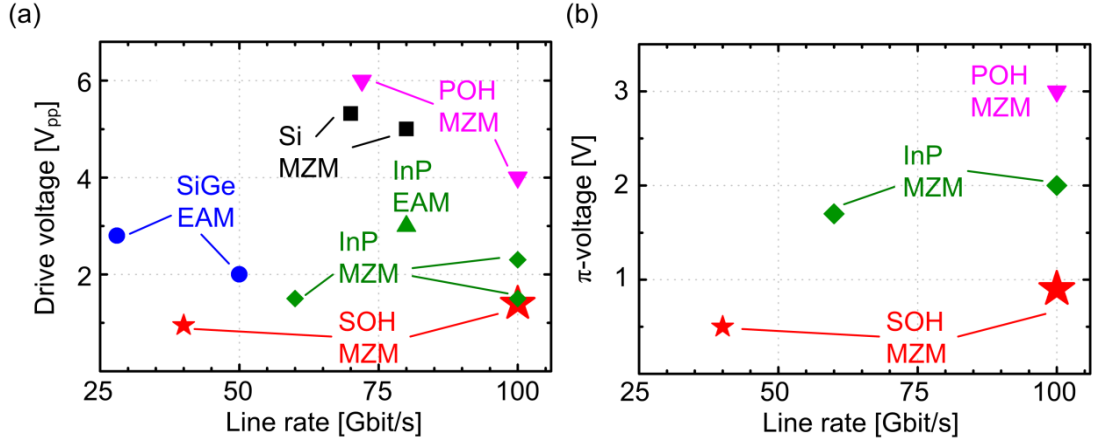

**Figure 5:** Comparison of OOK modulators fabricated on various material platforms. **(a)** Comparison of drive voltages and OOK line rates for different modulators types. Silicon germanium (**SiGe**) electro-absorption modulators (**EAM**), 28 Gbit/s @ 2.8 $V_{pp}$, 50 Gbit/s @ 2 $V_{pp}$ [12,13]. Indium phosphide (**InP**) EAM, 80 Gbit/s @ 3 $V_{pp}$ [8]. **InP MZM**, 60 Gbit/s @ 1.5 $V_{pp}$, 100 Gbit/s @ 2.3 $V_{pp}$ [9,10] and @ 1.5 $V_{pp}$ [11]. All-silicon Mach-Zehnder modulators (**Si MZM**) 70 Gbit/s @ 5.3 $V_{pp}$, 80 Gbit/s @ 5 $V_{pp}$ [14,15]. Plasmonic-organic hybrid (**POH**) MZM, 72 Gbit/s @ 6 $V_{pp}$ [36] and 100 Gbit/s @ 4 $V_{pp}$ [16]. Present work: Silicon-organic hybrid (**SOH**) MZM, 100 Gbit/s @ 1.4 $V_{pp}$, corresponding to a switching energy of 98 fJ/bit. This corresponds to the by far lowest value demonstrated to date. Another experiment with SOH modulators [22] was performed with drive voltages of 950 mV at a line rate of 40 Gbit/s. **(b)** Comparison of modulators in terms of OOK line rate and π-voltage. The **SOH MZM** for 40 Gbit/s and 100 Gbit/s exhibit π-voltages of 0.5 V and 0.9 V, respectively. These values are clearly below those achieved by other material platforms: **InP MZM** with π-voltages of 1.7 V and 2 V were operated at 60 Gbit/s and 100 Gbit/s, respectively. For the **POH MZM**, we estimate a $V_\pi$ of 3 V. In the plot, EAM are not depicted since the π-voltage is not a relevant figure of merit for these devices. For the **Si MZM**, the π-voltage is either not given or π-voltage of more than 20 V are estimated [14], which are outside the plotted parameter range. Similarly, the π-voltage of approximately 10 V in the 72 Gbit/s POH experiment cannot be depicted.

# Silicon-Organic Hybrid (SOH) Mach-Zehnder Modulators for 100 Gbit/s On-Off Keying
## - Supplementary Information -

Stefan Wolf[1†], Heiner Zwickel[1†], Wladislaw Hartmann[1,2,3†],
Matthias Lauermann[1,4], Yasar Kutuvantavida[1,2], Clemens Kieninger[1,2],
Lars Altenhain[5], Rolf Schmid[5], Jingdong Luo[6], Alex K.-Y. Jen[6], Sebastian Randel[1],
Wolfgang Freude[1], and Christian Koos[1,2,*]

[1]**Institute of Photonics and Quantum Electronics (IPQ), Karlsruhe Institute of Technology (KIT), 76131 Karlsruhe, Germany**
[2]**Institute of Microstructure Technology (IMT), Karlsruhe Institute of Technology (KIT), 76131 Karlsruhe, Germany**
[3]**Now with: Physikalisches Institut, University of Muenster, 48149 Muenster, Germany**
[4]**Now with: Infinera Corporation, Sunnyvale, CA, United States**
[5]**Micram Microelectronic GmbH, 44801 Bochum, Germany**
[6]**Department of Material Science and Engineering, University of Washington, Seattle, Washington 98195, USA**

†**these authors contributed equally to the work**
*email: christian.koos@kit.edu*

**Chirp analysis of SOH modulators**

For transmission of intensity-modulated signals, an unwanted phase modulation, also referred to as chirp, leads to increased sensitivity with respect to fiber dispersion and hence leads to a signal quality penalty[1]. Ideally, SOH MZM should not exhibit any chirp – the field-induced refractive index change of the polymer cladding according to the Pockels effect[2] allows to completely suppress chirp in an ideal, perfectly balanced push-pull Mach-Zehnder modulator (MZM)[3]. For real-world devices with finite extinction ratio (ER), however, chirp is introduced by an unbalance of the optical amplitudes in the two MZM arms. For quantifying the chirp of a data signal generated by such a device, we use the chirp parameter $\alpha$ that is essentially defined by the ratio of the phase modulation to the amplitude modulation[4,5], and that can be estimated from a direct measurement of the static extinction ratio $\delta^{(stat)}$, see Equations (6) and (7) of the main manuscript. To proof that finite static extinction ratio (ER) of the MZM is the dominant source of chirp in our devices, we compare the chirp parameter obtained from the ER to a direct measurement of $\alpha$. To this end, we use another SOH modulator with a device layout similar to that of the device used for the data transmission experiments, but with better static extinction ratio of $\delta^{(stat)} = 31$ dB.

For a direct measurement of the chirp parameter, we rely on the fiber response peak method[6]. Using a network analyzer, we measure the frequency response of the modulator with a 75 km fiber span. Erbium-doped fiber amplifiers (EDFA) are used to compensate the insertion loss of the modulator and the fiber span. The measured frequency response normalized to the back-to-back transmission is shown in Figure 1 a). For data evaluation, the product of the square of the resonance frequency $f_\mu$ and the fiber length $L$ is plotted in dependence of the resonance order $\mu$, see Figure 1 b). The chirp parameter $\alpha$ and the fiber dispersion $D$ can then be extracted from the slope and from the vertical offset of a straight line fitted to the data points[6]. Measurement analysis leads to a dispersion coefficient of 16.8 ps/(nm km) for the 75 km fiber span and chirp factors $\alpha$ of 0.09. The value compares well to the value $|\alpha| = 0.06$ which we estimate directly from the measured (static) extinction ratio. This confirms that the chirp of SOH MZM is dominated by the influence of a finite ER – the slight differences are attributed to the finite measurement accuracies of the fiber response peak method.

Using the equations described above, we estimate a chirp factor $|\alpha| \approx 0.42$ for a static extinction ratio of 14 dB as obtained for the MZM used for the 100 Gbit/s data experiment in the manuscript, see discussion in the main manuscript.

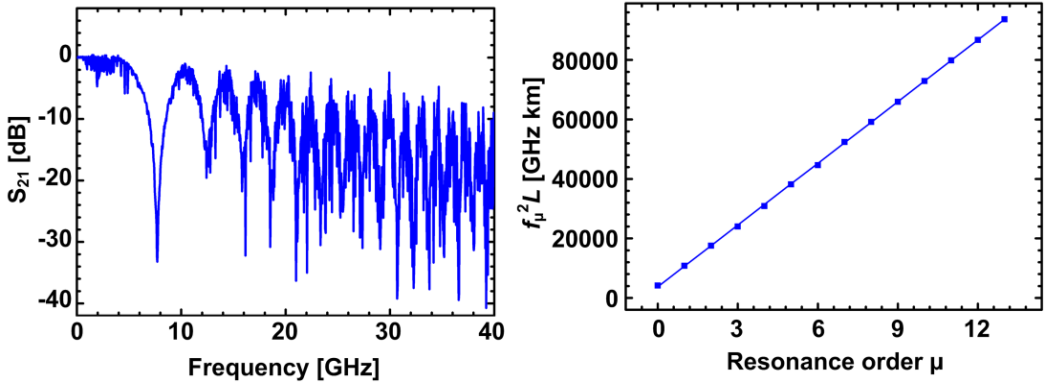

**Figure 1:** Chirp characterization of SOH Mach-Zehnder modulator. (a) Frequency response of SOH MZM with 75 km fiber span for modulator operated without gate field. The frequency response has been normalized to the modulator's back-to-back characterization. Resonant dips originate from fiber dispersion. (b) From the product of the squared resonance frequency and the fiber length, $f_\mu^2 L$, fiber dispersion parameter and chirp factor $\alpha$ can be estimated[6]. In the analysis, we find a chirp factor of $\alpha \approx 0.09$.

## Wavelength operating range of SOH modulators

SOH modulators can operate over a wide range of wavelengths. In essence, the operating range is only limited by the ability of the silicon photonic slot waveguide to efficiently guide the light and by the absorption of the organic EO material. The modulator structure as well as the organic material are well suited to operate over the entire range of infrared telecommunication wavelengths, comprising all relevant transmission bands between 1260 nm and 1675 nm. Absorption of the EO material is negligible above 1200 nm [7,8] and hence the material is transparent in wavelength regions beyond 1260 nm which are relevant for telecommunications. The performance of the slot waveguides can be quantified by considering the field interaction factor according to Eq. (2) of the main manuscript. For the device used in the transmission experiments of the main manuscript, the field interaction factor decreases from $\Gamma \approx 0.16$ at $\lambda = 1550$ nm to $\Gamma \approx 0.14$ at $\lambda = 1310$ nm, see Figure 2 for the associated mode field profiles. At the same time, the material-related figure-of-merit for EO activity, $n_{EO}^3 r_{33}$, increases[9] by about 40 %, which leads to an overall improvement of the modulation efficiency by about 16 %. In total, the device performance does hence not change significantly with wavelength.

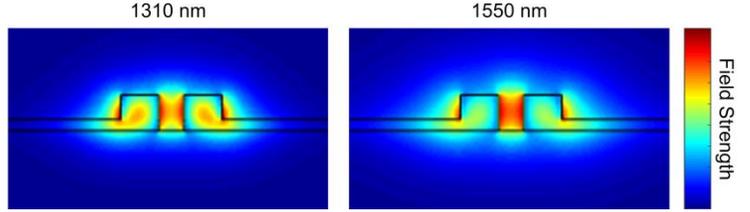

**Figure 2:** Simulated mode fields for a center wavelength of 1310 nm (left) and 1550 nm (right). The color coding indicates the modulus of the transverse electrical field and is identical for both plots. For the shorter wavelength, the field is stronger in the silicon rails, and hence the field interaction factor is smaller by 17 %. This is overcompensated by an increase of the material-related figure-of-merit for EO activity, $n_{EO}^3 r_{33}$.

## Implementation of pulse shape for system simulation

The hardware used in our experiment features nonzero rise and fall times, see eye diagram in Figure 3 (a) for the measured electrical output signal. For an emulation of these signals in the simulation, we use cosine-shaped pulses in the time-domain (not to be confused with raised-cosine pulse shaping with a raised-cosine shaped spectrum). A mathematical description of the pulse shape $p(t)$ is given as

$$p(t) = \frac{1}{2}\cos\left(\pi \frac{t}{T}\right), \qquad -T \leq t \leq T, \tag{1}$$

where $T$ denotes the symbol duration. With the discrete symbols $a_k = \{-1;1\}$ which correspond to a logical "1" and a logical "0", respectively, with the Dirac function $\delta(t)$ representing the time-discrete nature of digital data, and with the peak-to-peak output voltage $U_{pp}$, the (electrical) mean-free time-domain drive signal can be written as

$$s(t) = U_{pp} \sum_k a_k \delta(t - kT) \otimes p(t) \tag{2}$$

where $\otimes$ denotes a convolution operation. The eye diagram of the digitally generated waveform (with additive white Gaussian noise) is depicted in Figure 3 (b).

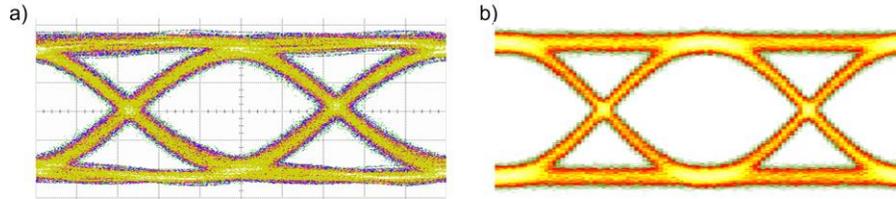

**Figure 3: Waveform for bandwidth analysis.** (a) Measured eye diagram from the MUX at 100 Gbit/s, having a peak-to-peak voltage swing of approximately 0.4 $V_{pp}$. The signal is fed to an RF amplifier and to the SOH modulator. (b) Eye diagram of digital waveform that is used to approximate the output signal of the MUX for the bandwidth simulation in the main manuscript.